\documentclass[12pt,a4paper,final]{iopart}

\usepackage{iopams}  
\usepackage[breaklinks=true,colorlinks=true,linkcolor=blue,urlcolor=blue,citecolor=blue]{hyperref}
\usepackage[dvips]{graphicx}
\usepackage{bm}

\usepackage{color,ulem}

\newcommand{\blue}[1]{\textcolor[rgb]{0,0,1}{#1}}

\begin{document}
\title{Reliability of the one-crossing approximation in describing the Mott transition}

\author{V. Vildosola}
\address{Depto de F\'{i}sica CAC-CNEA and Consejo Nacional de Investigaciones Cient\'{\i}ficas
 y T\'ecnicas, CONICET, Rep\'{u}blica Argentina}
 
\author{L. V. Pourovskii}
\address{Centre de Physique Th\'{e}orique, \'{E}cole Polytechnique, CNRS, 91128 Palaiseau, France}
 
\author{L. O. Manuel}
\address{Instituto de F\'{i}sica Rosario, Consejo Nacional de Investigaciones 
Cient\'{i}ficas y T\'{e}cnicas
and Universidad Nacional de Rosario, Bvd. 27 de Febrero 210 Bis, 2000 Rosario, 
Rep\'{u}blica Argentina}

\author{P. Roura-Bas}
\address{Depto de F\'{i}sica CAC-CNEA and Consejo Nacional de Investigaciones Cient\'{\i}ficas
 y T\'ecnicas, CONICET, Rep\'{u}blica Argentina}

\begin{abstract}

We assess the reliability of the one-crossing approximation (OCA) approach in
quantitative description of the Mott transition in the framework of the dynamical
mean field theory (DMFT). The OCA approach has been applied in the conjunction with
DMFT to a number of heavy-fermion, actinide, transition metal compounds, and
nanoscale systems.  However, several recent studies in the framework of impurity
models pointed out to serious deficiencies of OCA and raised questions regarding its
reliability. Here we consider a single band Hubbard model on the Bethe lattice at
finite temperatures and compare the results of OCA to those of a numerically exact
quantum Monte Carlo (QMC) method. The temperature-local repulsion $U$ phase diagram
for the particle-hole symmetric case obtained by OCA is in good agreement with that
of QMC, with the metal-insulator transition captured very well. We find, however,
that the insulator to metal transition is shifted to higher values of $U$ and,
simultaneously, correlations in the metallic phase are significantly overestimated. 
This counter-intuitive behavior is due to simultaneous underestimations of the Kondo
scale in the metallic phase and the size of the insulating gap. We trace the
underestimation of the insulating gap to that of the second moment of the
high-frequency expansion of the impurity spectral density.  Calculations for the
system away from the particle-hole symmetric case are also presented and discussed.

\end{abstract}


\maketitle

\section{Introduction}

In the past years, many efforts have been devoted to the implementation of calculation techniques to describe 
the electronic structure of strongly correlated complex materials. This is a complicated and challenging task 
in view of the many degrees of freedom involved. One of the most successful approaches in this direction 
was the implementation of the dynamical-mean field theory (DMFT) \cite{dmft-1, dmft-2, dmft-3}. Numerically, the most challenging part of DMFT
is the solution of the Anderson impurity model \cite{anderson} within the  DMFT self-consistent loop 
that maps the lattice problem into a single impurity one. 

There are two well-known numerically exact techniques to solve this impurity model, namely, the quantum Monte Carlo (QMC) 
in its Hirsch-Fye (HF-QMC) or continuous time (CT-QMC) versions \cite{hirsch, ct-qmc}, and 
the numerical renormalization group (NRG) \cite{nrg,nrg-dmft}. 
Recently, a substantial technical progress \cite{andreas-1} has been achieved in both approaches. 
On one hand, the advent of continuous-time quantum Monte Carlo methods \cite{Gull_QMC_review} eliminated 
the time discretization error, inherent to the HF-QMC, and extended the range of applicability of QMC to 
much lower temperatures and realistic Coulomb repulsion vertices. On the other hand, very fast 
implementations of NRG applied to multi-band systems has been developped using Abelian and non-Abelian 
symmetries on a generic level \cite{andreas-2}.

In spite of recent technical improvements, those exact methods still encounter certain difficulties.
QMC solvers suffer from the well known 'fermion sign problem', which can be especially severe when the degeneracy 
of the correlated shell is large and significant off-diagonal terms are present in the hybridization function. 
Moreover, QMC calculations are carried out in the imaginary-time domain and an analytic continuation is required to 
obtain real-energy spectral functions from QMC data.
The NRG approach becomes computationally expensive  
in multiorbital cases with broken orbital symmetries (for instance, when  
interactions, like pair-hopping, prohibit the use of 
symmetries that reduce the size of the matrix to be diagonalized \cite{pruschke-bulla}, leading to an exponential
increase of the Hilbert space).  Because of these limitations the necessity to have faster and reliable impurity solvers 
is evident. 

Hence, several approximate schemes have been
proposed for solving the DMFT impurity problem, like the 
local moment approximation (LMA)\cite{lma},
iterative perturbation theory (IPT)\cite{ipt}, exact diagonalization\cite{ed},
rotationally invariant slave bosons \cite{lechermann}, 
conserving diagrammatic approximations based on self-consistent hybridization expansion (SCH) \cite{conserving}, 
among others. 

Regarding the SCH, the non-crossing approximation (NCA) \cite{nca} represents 
the simplest family of these self consistent treatments and provides an accurate calculation
of the impurity Green function, as well as many other properties, when the Coulomb repulsion is taken
to be large enough as compared with the other energy scales involved in the problem. 
However, when more than one charge fluctuation needs to be included ($N\rightarrow N-1$ and $N\rightarrow N+1$, 
being $N$ the impurity valence), NCA has failed to give the correct Kondo scale ($T_K$).
The next leading order in the self consistent expansion, that partially solves this pathology, is often
known as the one-crossing approximation, OCA \cite{oca-1,oca-2,oca-3}. Within this extended formalism 
other classes of problems have been investigated \cite{haule-2, haule-1, schmitt-1}. Among them, its major application is in 
the context of the dynamical mean field theory as an impurity solver \cite{dmft-3}.

In particular, the OCA solver has the advantage of being formulated at the real frequency axis and 
it gives the correct order of magnitude for the Kondo scale of the impurity problem. 
It successfully captures the correct temperature dependence of transport properties 
of a single impurity level \cite{oca-3}, and it has been employed as the DMFT impurity 
solver in a search for signatures of a non-Fermi liquid 
behavior in the Hubbard model with van Hove singularities \cite{schmitt-1}.  
Furthermore, it has been generalized to an arbitrary number of orbitals and interactions \cite{haule-1}.
Multiorbital generalization of OCA were employed in a study of
the itinerant and local-moment magnetism in the three-band Hubbard model \cite{schmitt-2}. 
In combination with \textit{ab-initio}$+$DMFT calculations, the OCA solver has been applied to real strongly correlated 
materials, for example, to heavy-fermion compounds \cite{dmft-3, haule-1, haule-1-2, haule-2}.

However, the OCA solver has also several limitations. It cannot be applied to arbitrary low temperatures
due to violations of the Fermi-liquid properties (in the impurity model, OCA works well for $T > 0.1 T_K$) \cite{oca-3,grewe-1,grewe-2,grewe-3},  
and it also violates the sum rules for the coefficients of the high-frequency expansion of the self-energy
\cite{millis}. While the former pathology can be controlled by restricting its application to high enough 
temperatures, the later one is intrinsic and will always be present.
As has been pointed out recently, the OCA method is more accurate in the 
strongly-correlated limit \cite{millis}, and it describes the insulating phase particularly well \cite{ruegg}.
It has also been shown that OCA overestimates the correlations in th metallic phase and it has been conjectured that 
this overestimation of correlation effects reflects the fact that the OCA
tends to favor the insulating state. 

One important issue that has not been 
studied up to date is the actual quantitative performance  of the 
OCA solver within DMFT in describing the metal-insulator Mott transition\cite{mott}. 
Hence, we address this issue in the present work by calculating the  critical $U_c$ values for the Mott transition  within DMFT as a function of 
temperature using OCA as the impurity solver,  
and comparing them with the corresponding ones obtained with the CT-QMC. We have also compared the DMFT local self-energies obtained within the two approaches as well as the corresponding quasi-particle effective masses in the metallic phase. Our calculations have been carried out for the 
single band Hubbard model with a semicircular non-interacting density of states.

Our main conclusion is that the OCA 
metal-to-insulator transition for the particle-hole symmetric case is in remarkably good agreement with  
that of CT-QMC. However, we find that insulator-to-metal transition is shifted to higher values of $U$ despite the fact 
the correlations of the metallic phase are overestimated. This counter-intuitive behaviour is explained as a combination of two factors: 
the underestimation of the effective Kondo temperature in the metallic phase and the underestimation of the gap in the insulating one. 
The fact that OCA underestimates the gap in the insulating regime comes out from an analysis of the high-frequncy expansion 
sum rules of the Green function. Our results are in contradiction to the conjecture of OCA favoring the insulating phase. 
We show that although OCA overestimates the strength of correlations in the metallic phase, it does not favor the insulating 
one because the critical values of the metal-to-insulator transition are very well captured.

We have also study the same model in the non-symmetric case, obtaining similar agreement between both 
techniques. We verify that the OCA approximation does not violate 
the Friedel sum rule in the metallic phase for the range 
of temperatures of the obtained phase diagram, and that the interacting part of the OCA self-energy always remains 
causal. 

The paper is organized as follows: we describe the theoretical formalism in section \ref{model}, we present the numerical results for the 
particle-hole symmetric case in section \ref{symm}, we discuss the results obtained for the system away from half-filling in section \ref{non-symm}
and finally we conclude in section \ref{conclusions}.

\section{Model and Formalism}\label{model}
\label{modelo}

We start with the single-band Hubbard Hamiltonian,

\begin{eqnarray}\label{Hubbard}
H= -\frac{t}{\sqrt{z}}\sum_{<ij>\sigma}( c^{\dagger}_{i\sigma}c_{j\sigma} +
                                         c^{\dagger}_{j\sigma}c_{i\sigma} ) + 
   U\sum_{i} n_{i\uparrow}n_{i\downarrow}, 
\end{eqnarray}
where the first term is the kinetic energy, $t$ is the hopping between nearest neighbors on a lattice, $z$ is 
the coordination number, and $U$ is the energy of the on-site Coulomb 
repulsion. The operator
$c^{\dagger}_{i\sigma}$ creates an electron with the spin $\sigma$ on the  site $i$ and 
$n_{i\sigma}=c^{\dagger}_{i\sigma}c_{i\sigma}$. We use the semicircular non-interacting density of 
states  $N(\omega)=\frac{1}{2\pi t^2}\sqrt{4t^2-\omega^²}$, 
$\vert\omega\vert<2t$ corresponding to a Bethe lattice with coordination $z \rightarrow\infty$, 
for which the DMFT approximation becomes exact. In the following we use the half bandwidth as our
unit of energy $D=2t=1$.

We solve the Hamiltonian [\ref{Hubbard}] by means of DMFT, which maps the lattice model 
onto a single-impurity Anderson one within a self-consistent cycle. The hopping
between the impurity and the conduction band, $V_k$, defines the hybridization function for the single-impurity problem
$\Gamma(i\omega)=\sum_{k}V_k^2 / (i\omega-\epsilon_k)$, where $\epsilon_k$ is the conduction
energy of the impurity model. Within the DMFT and in the case of the Bethe lattice, the DMFT hybridization function is 
given by the self-consistency condition $\Gamma(i\omega)=t^2G[\Gamma(i\omega)]$, where $G(\omega)$ is the local 
Green function obtained from the impurity model.

Starting from the metallic non-interacting solution of the model, the system turns into an insulator for large enough 
values of the Coulomb repulsion $U$ due to the vanishing of the quasiparticle weight. 
The value of $U=U_{c2}$ defines this transition. On the other hand, starting from an insulating solution, the systems 
turns metallic due to the collapse of the gap between the Hubbard bands, 
for $U \le U_{c1}$,  with $U_{c1} < U_{c2}$ when $T$ is lower than the second-order end point of the first-order Mott 
transition $T_c$. 
The critical values $U_{c1}<U<U_{c2}$ as function of the temperature $T$ determine a phase diagram.

The phase diagram of the Mott transition for the present model have been previously obtained
using the QMC \cite{rozenberg,oudovenko,blumer}, IPT \cite{dmft-1}, exact diagonalization \cite{caffarel,rozenberg94,tong}, and 
NRG\cite{nrg-dmft} impurity solvers. The determination of the exact boundaries of the coexistence region has previously required a 
significant effort due to their sensitivity to calculational parameters, as well as due to the critical slowing down of the DMFT 
convergence close to those boundaries \cite{oudovenko}. Hence, we have employed up to 220 DMFT cycles for each point 
in the $\{U,T\}$ space and used a dense mesh along the $U$ axis with the spacing between $U$ values down to 0.005 in the 
vicinity of the $U_{c1}$ line. 
We have used the CT-QMC implementation provided by the TRIQS package \cite{triqs,triqs_paper}. The DMFT impurity problem has been solved by 
CT-QMC using $\sim 10^9$ CT-QMC moves with each 200 moves followed by a measurement. 
The resulting CT-QMC phase diagram is in agreement with the extensive HF-QMC calculations of Bl\"umer \cite{blumer}.
Within the OCA solver we have used the procedure described by Hettler \textit{et al.} for regularizing the 
spectral functions \cite{oca-reg} and the numerical convolution sketched in Ref.\cite{haule-2} when computing the
self-energies and the Green function.

\section{Numerical Results}\label{results}

In this section, we present the numerical results obtained using the OCA solver for the DMFT loop 
and a detailed comparison with CT-QMC calculations.
 
\subsection{Mott transition for the particle-hole symmetric case}\label{symm}

In order to get the critical values $U_{c1}(T)$ and $U_{c2}(T)$ for a given temperature $T$ 
within the OCA solver, we take advantage of its self-consistent nature building an external 
loop running in the $U$ values. Starting from a metallic solution we slowly increase $U$ by 
$\delta U$ retaining the previous ionic self-energies and Green function as the initial guess
for the following $U+\delta U$ DMFT cycle, until an insulator solution is reached, and then
we decrease $U$ by $-\delta U$ until we go back to the initial $U$. 

\begin{figure}[h!]
\includegraphics[clip,width=7cm]{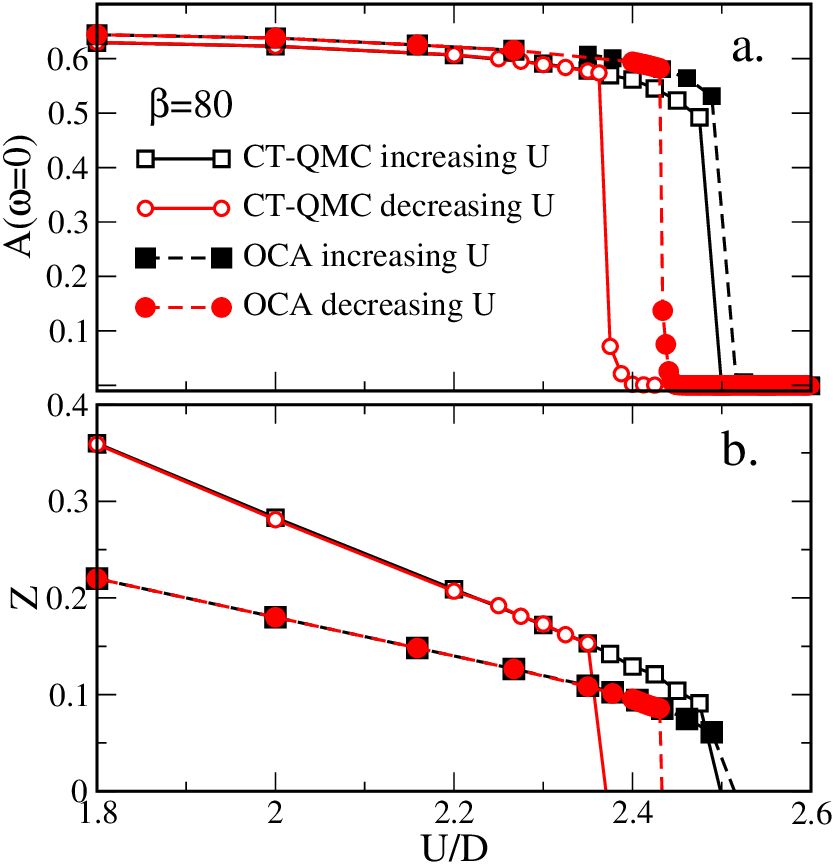}
\caption{(Color online) a). Spectral weight $A(\omega=0)$ for the inverse 
temperature $\beta=80$ as a function of $U$ both for  increasing (black lines, squares) and decreasing  (red lines, circles) $U$ values. 
The CT-QMC (OCA) data are displayed with the solid (dashed) lines and empty (filled) symbols, respectively. b). 
The quasi-particle residue $Z$ as function of $U$ for the same temperature. The notation is the same as in panel a). }
\label{Fig1}
\end{figure}

In Fig.[\ref{Fig1}\blue{a}] we show the spectral weight at the zero-frequency,
$A(\omega=0)=-\mathcal{I}m[G(\omega=0)]/\pi$, as
a function of $U$ for an inverse temperature $\beta=80$. We show both the increasing $U$ results
from the metallic to insulator solutions as well as the decreasing ones. An hysteresis curve 
is formed, giving rise to two different critical values, $U_{c1}(T)$ and $U_{c2}(T)$. We define these critical
values following the criteria given in Ref.\cite{nrg-dmft}, from the $U$-value for which 
$\vert A'(\omega=0)\vert$ reaches its maximum intensity. 

In Fig.[\ref{Fig1}\blue{b}] we show the variation of the quasiparticle weights, 
$Z=[1-\frac{\partial Re \Sigma(\omega)}{\partial\omega}\vert_{\omega=0}]^{-1}$, as a function of $U$
for $\beta=80$. 
In order to compared with CT-QMC, we first
obtain the interacting part of the OCA self-energy $\Sigma(\omega)$, removing the non-interacting 
offset given by the hybridization term. Secondly, from a Hilbert transform of $Im\Sigma(\omega)$, we 
compute the corresponding self-energy in the Matsubara domain,

\begin{equation}
 \Sigma(i\omega_n) = -\frac{1}{\pi}\int d\omega~ \frac{Im\Sigma(\omega)}{i\omega_n-\omega}.
\end{equation}

Finally, we approximate the derivative 
$\frac{\partial Re \Sigma(\omega)}{\partial\omega}\vert_{\omega=0}=
\frac{\partial Im \Sigma(i\omega_n)}{\partial i\omega_n}\vert_{i\omega_n\rightarrow0}$
by a cubic fitting of the first four Matsubara's frequencies of $Im \Sigma(i\omega_n)$.

Although the vanishing of $Z$ defines the critical value $U_c$ only at zero temperature \cite{nrg-dmft},
it has been used as a common criteria even for finite temperatures (see for instance Ref.\cite{ansgar})
From Fig.[\ref{Fig1}] it can be seen that both approaches (from $A(\omega=0)$ or 
from $Z$) define 
the same energy scales for $U_{c1}$ and $U_{c2}$. More importantly, the OCA critical $U$-values 
are in a reasonable agreement with the CT-QMC ones. 
While the OCA value for $U_{c2}$ is obtained within an error of less than 0.5\%  with respect to the CT-QMC one, 
the calculated $U_{c1}$ is larger than the CT-QMC one by around 3\% .  
We will discuss the origin of this discrepancy for $U_{c1}$ later in this section.

It is important to remark that the OCA values of $Z$ in the metallic region, i.e.  $U<U_{c1},U_{c2}$, are smaller than the 
CT-QMC ones. The same behavior was found by Schmitt \textit{et al.} \cite{schmitt-2} using OCA for
a body-centered-cubic 
lattice in comparison with NRG calculations. While OCA gives the correct low energy scale for the impurity
model, this energy scale is still slightly underestimated \cite{oca-2}, and therefore within OCA 
the system feels a larger effective Coulomb repulsion giving rise to a reduced quasiparticle weight. 
However, it is important to remark that the underestimation of Z is less important close to the transition.

\begin{figure}[h!]
\includegraphics[clip,width=7cm]{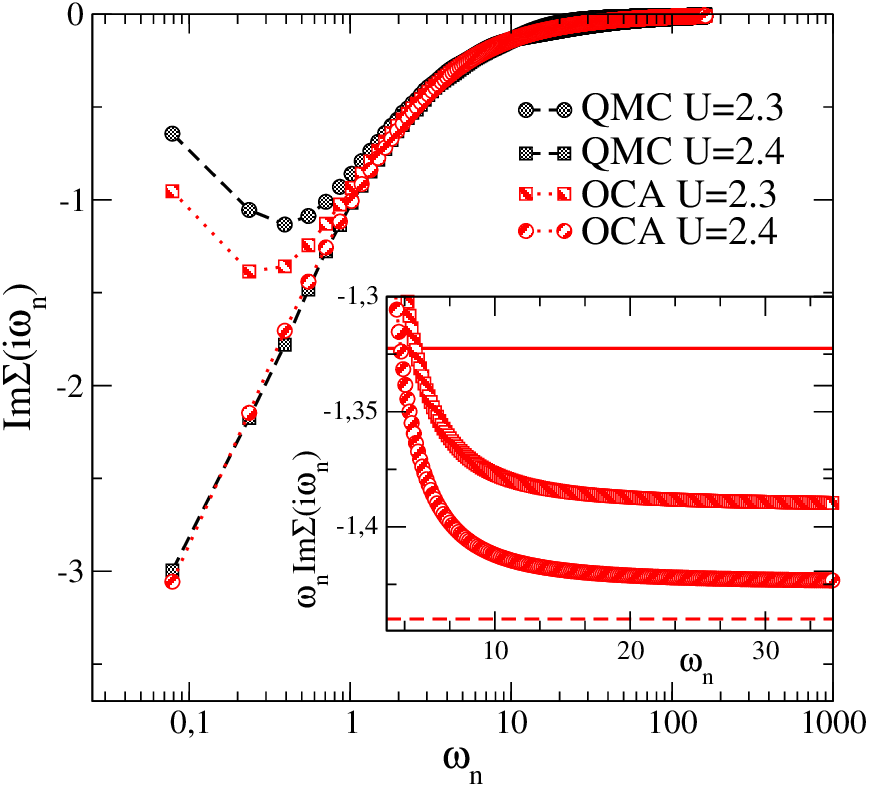}
\caption{(Color online) Comparison of the imaginary part of the 
self-energy as a function of the 
Matsubara frequency between OCA and CT-QMC at  $\beta=60$ for two different values of $U$, one below the 
$U_{c2}$ and the other one above. The inset shows the imaginary part of 
the OCA self-energy scaled by $\omega_n$. The dashed and solid
lines indicate their expected theoretical values given by the high frequency expansion sum rule, 
$\Sigma_1 = -U^2/4$.}
\label{Fig2.eps}
\end{figure}

In Fig.(\ref{Fig2.eps}) we show the imaginary part of the self-energy in the imaginary 
frequency domain for the increasing $U$ regime at  $\beta=60$ and for two different values of $U$, 
one below and one above $U_{c_2}$, $U=2.3$ and $U=2.4$. 
As it can be observed from this plot, for the metallic case, OCA overestimates the absolute magnitude of the 
self-energy at low frequencies. 
Similarly to the underestimation of the quasiparticle weigth at low temperatures described above, this behavior of 
$Im\Sigma(i\omega_n)$  can be also understood as arising due to an effectively larger value of $U$. On the other 
hand, in the insulating
region the agreement between OCA and CT-QMC is remarkable. We found that for a correct comparison between the two 
techniques it was very important to have the same degree of precision of the convergence criterion of the DMFT 
loops, especially for points close to the Mott transtion. 
For large frequencies, an additional test can be done using the sum rules that 
$\Sigma(i\omega_n)$ should satisfy.

In the inset of Fig.(\ref{Fig2.eps}) we plot the imaginary part of 
the OCA self-energy scaled by $\omega_n$ for $U=2.3 < U_{c2}$ and $U=2.4 > U_{c2}$, 
together with the exact coefficient $\Sigma_1$ for each $U$, that corresponds 
to the first moment in the self-energy high frequency expansion, 
$\Sigma_1=\int\frac{d\omega}{\pi}~Im\Sigma(\omega),$ and that 
determines the asymptotic $1/\omega_n$ behavior. In Ref.\cite{millis},
R\"{u}egg \textit{et al.} have calculated the exact value expected for $\Sigma_1$, being 
$\Sigma_1=-U^2/4$ for the symmetric case \cite{sigma-1}. For the parameters shown in
Fig.(\ref{Fig2.eps}), we obtain a deviation of 
the OCA $\Sigma_1$ coefficient of the order of $5\%$ in the metallic phase,
while in the insulator one the error is reduced to less than $2\%$.

\begin{figure}[h!]
\includegraphics[clip,width=7cm]{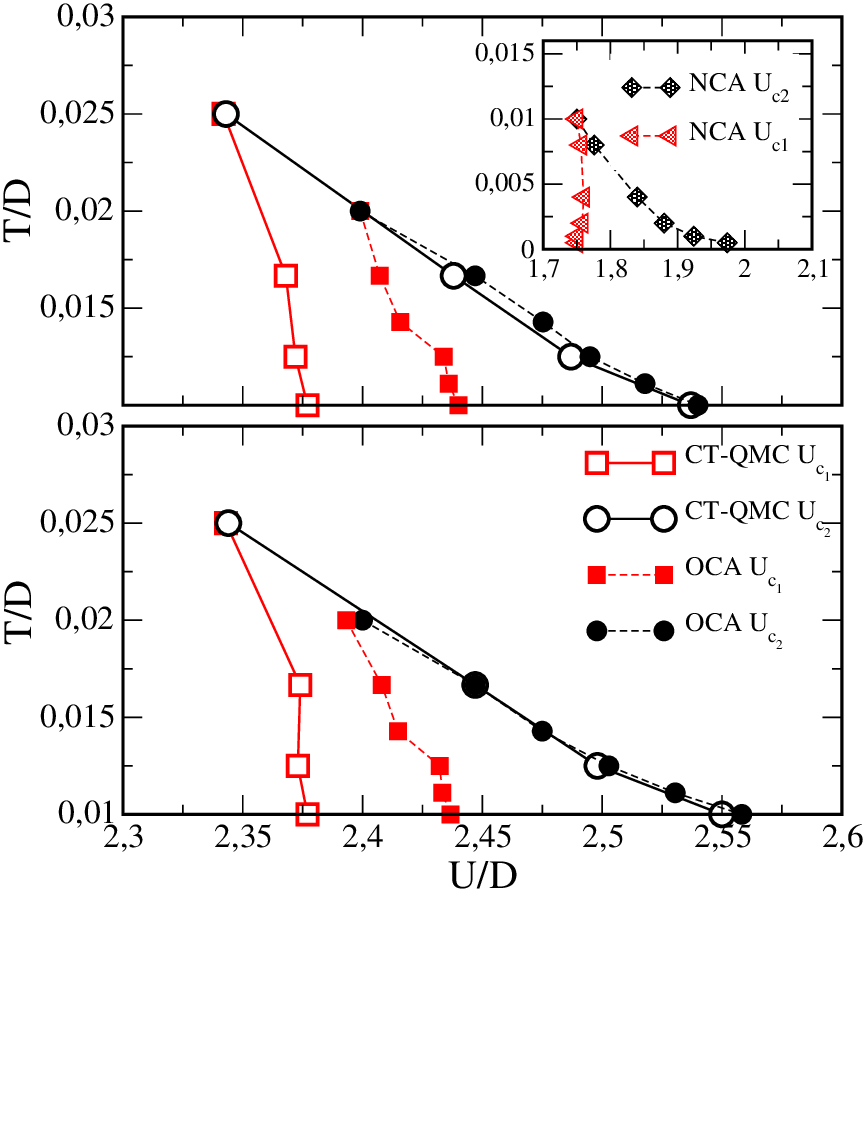}
\caption{(Color online) The $T$ vs. $U$  phase diagram of the Mott transition obtained from the 
zero-frequency spectral function $A(\omega=0)$ (upper panel) and the quasiparticle residue $Z$ (lower panel).
The inset in the upper panel indicates the phase diagram obtained using the finite-U NCA in the symmetric case as the impurity solver.}
\label{Fig3}
\end{figure}

In what follows we discuss the phase diagram of the Mott transition.
In Fig.[\ref{Fig3}] we show the $T$ vs. $U$ diagram with the calculated  $U_{c_1}$ and $U_{c_2}$ obtained from the 
zero-frequency spectral function $A(\omega=0)$ (upper panel), as well as the quasiparticle residue $Z$ (lower panel).
The general trend of the critical $U_c(T)$ obtained by OCA is in reasonable agreement with the corresponding
CT-QMC ones.
Even though a very well defined coexistence region is captured by OCA,  
this coexistence region is reduced with respect to the 
CT-QMC one. 
While the agreement is remarkable for the $U_{c2}(T)$ transition, the  $U_{c1}(T)$ values are slightly shifted to higher energies in OCA. 

Regarding the critical temperature ($T_c$) below which two different spinodal lines define the 
coexistence region of the insulating and metalic regimes of the Mott transition, OCA gives $T_c\sim 0.02$ 
in reasonable agreement with the CT-QMC $T_c\sim0.025$.
The slight underestimation of $T_c$ is a consequence of the corresponding underestimation of $T_K$ by OCA at the effective impurity level. 
For comparison, we also include in the inset of the upper panel of Fig.[\ref{Fig3}] the finite-U NCA phase diagram for the particle-hole symmetric case. 
We stress here that this simple approximation severely underestimates all the energy scales involved, 
$T_c$ and both $U_{c1}(T)$ and $U_{c2}(T),$ as a consequence of the underestimated Kondo scale.
On the other hand, we want to mention here that the IPT results \cite{dmft-1, nrg-dmft} are considerably shifted to  
higher energies overestimating both, $U_{c1}(T)$ and $U_{c2}(T)$, due to the exaggerated overestimation of the Kondo 
scale at the impurity level.

Despite its approximate nature, the coexistence region given by OCA is in the correct energy range
and the critical temperature $T_c$ is in very good agreement with the CT-QMC results.  
We want to remark that for the whole range of temperatures studied in the presented phase diagram,
the OCA self-energy remains causal, that is, $Im \Sigma(i\omega_n)$ is negative. For very low temperatures 
($T\lesssim 1/500 \sim 0.1 \;T_c^{OCA}$), it can turn positive signaling the breakdown of the approximation.

We turn now to the discussion regarding the slight overestimation of $U_{c1}$ that can be observed  in Fig.(\ref{Fig3}).
While the value of $U_{c2}$ is given by the critical $U$ for which the quasiparticle weight at 
zero frequency vanishes, the $U_{c1}$ is related to the corresponding 
$U$ for which the Hubbard bands collapse and the gap in the spectral function is closed. 
We found that the size of the gap in the insulator regime given by OCA is somewhat underestimated 
and therefore it closes for a larger value of $U$ than expected for CT-QMC. 
This statement follows from an analysis of the high frequency expansion of the local Green function.
As described in Ref.\cite{millis}, the high frequency expansion in the imaginary domain of $G(i\omega_n)$ is
given by

\begin{equation}
 G(i\omega_n) = \sum_{k=1}^{\infty} \frac{M_{k-1}}{(i\omega_n)^k},
\end{equation}
where, in the spectral representation of the Green function, the coefficients are related to the moments 
of the spectral density as $M_k= \int_{\infty}^{\infty}d\omega~\omega^kA(\omega)$ \footnote{With our notation the moments $M_k$ are equal to the coefficients $c_{k+1}$ defined in Ref.~\cite{millis}}  . Exact relations for the 
coefficients can be found from thermodynamic expectation values \cite{millis}: $M_0 = 1$, 
$M_1=\epsilon_d+Un_d/2$ ($0$ at half filling),
and $M_2=\epsilon_d^2+\Delta_1+U(2\epsilon_d+U)n_d/2$. Here, $\epsilon_d$ and $n_d$ are the energy level and total
occupancy of the effective Anderson model. $M_0$ and $M_1$ are related
to the normalization and parity of $A(\omega)$ so that they are exactly reproduced by OCA.  

Regarding the coefficient $M_2$, the parameter $\Delta_0$ represents the zero moment in the hybridization
high frequency expansion, $\Delta_0=-\frac{1}{\pi}\int_{\infty}^{\infty}d\omega~\mathcal{I}m\Gamma(\omega)
=\frac{1}{\pi}\int_{\infty}^{\infty}d\omega~\Delta(\omega)$, where $\Delta(\omega)=\pi V^2 \rho_c(\omega)$, 
and $\rho_c$ is the conduction density of states. Using the self-consistency condition 
$\Gamma(i\omega)=t^2G[\Gamma(i\omega)]$ for the present case of the Bethe lattice, we arrive to the following
relation: $\Delta(\omega)=\pi t^2 A(\omega)=\frac{\pi D^2}{4} A(\omega)$. Therefore,   
$\Delta_0=\frac{D^2}{4}\int_{\infty}^{\infty}d\omega~A(\omega)=\frac{D^2}{4}$. Taking into account that for 
the symmetric situation $2\epsilon_d+U=0$ and $M_1=0$,  the coefficient $M_2$ reads

\begin{equation}\label{c3}
 M_2 = \frac{U^2}{4} + \frac{D^2}{4}.
\end{equation}

\begin{figure}[h!]
\includegraphics[clip,width=7cm]{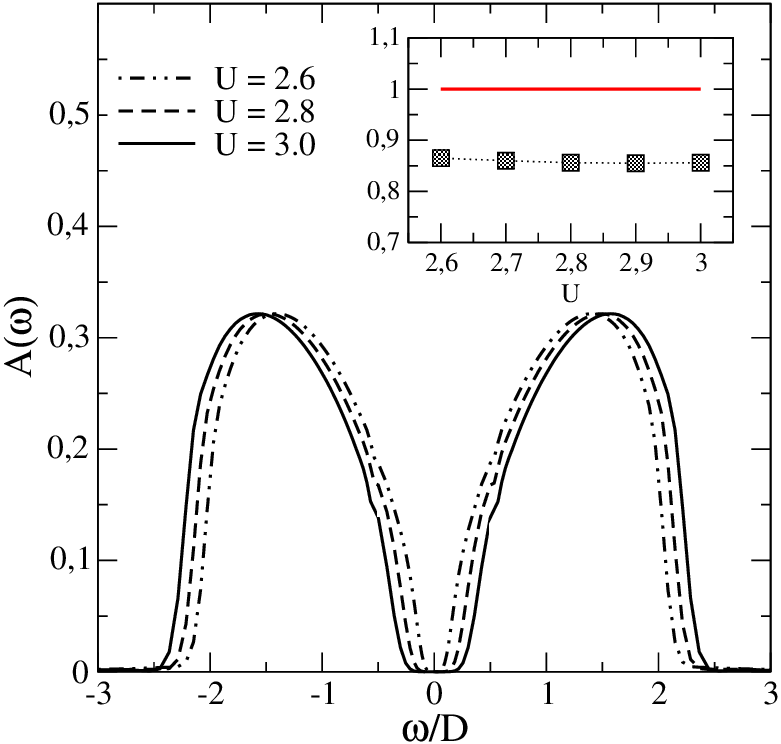}
\caption{(Color online) Spectral density in the insulating region when decreasing the Coulomb
repulsion from $U=3$ to $U=2.6$. The inset shows the ratio of the second moment obtained within 
OCA and its exact value from Eq.(\ref{c3}) (squares) as a function of $U$ and its deviation
from the unity (solid line). }
\label{Fig4}
\end{figure}

The second moment $M_2$ of the spectral function contains indirect information about the size of the Mott gap. 
In fact, it carries information about the center position and width of each Hubbard band. For instance, in the simplest
case in which the Hubbard bands have the semicircular shape centered at $\pm \omega_0$ and width $D$, the second 
moment becomes $M_2 = \omega_0^2 + D^2/4$ by comparing with Eq.(\ref{c3}), one can infer that $\omega_0=U/2$.
In this simple picture, the gap is opened when $U$ is larger than $2D$ and the size of the gap is 
of the order of $\delta=U-2D$. 
In Fig.(\ref{Fig4}), we show the spectral density in the insulating region when decreasing the Coulomb
repulsion from $U=3$ to $U=2.6$. It can be observed that the gap is continuosly closed when $U$ is lowered until  
the critical value $U_{c1}$ is reached. In the inset of Fig.(\ref{Fig4}), we show the values of 
$\frac{4}{U^2+D^2}\int_{\infty}^{\infty}d\omega~\omega^{2}A(\omega)$ (squares), which represents the ratio of 
the second moment obtained within OCA and its exact value from Eq.(\ref{c3}), as a function of $U$ and its deviation
from the unity (solid line). It can be seen that OCA underestimates the second moment of the spectral function by $\sim15\%$. 

Unfortunately, the center position and width of each Hubbard band enter in a combination within $M_2$ and we cannot
know from this coefficient alone, if OCA underestimated the center position or width or even both. 
However, an underestimation in both quantities bring about a reduccion of the gap that gives rise to larger values
of $U_{c1}$ as compared with the exact CT-QMC ones.

\subsection{Non-symmetric case}\label{non-symm}

In this subsection, we compare the calculations done by OCA and CT-QMC for the one band Hubbard model on the 
Bethe lattice away from half-filling. We consider  $2.5 < U < 5.0$ and the impurity level of the effective Anderson model 
at $\epsilon_d=-\frac{U}{2}+\Delta \mu$, with $\Delta\mu$=-1.0 and $\beta= 60$.

\begin{figure}[h!]
\includegraphics[clip,width=7cm]{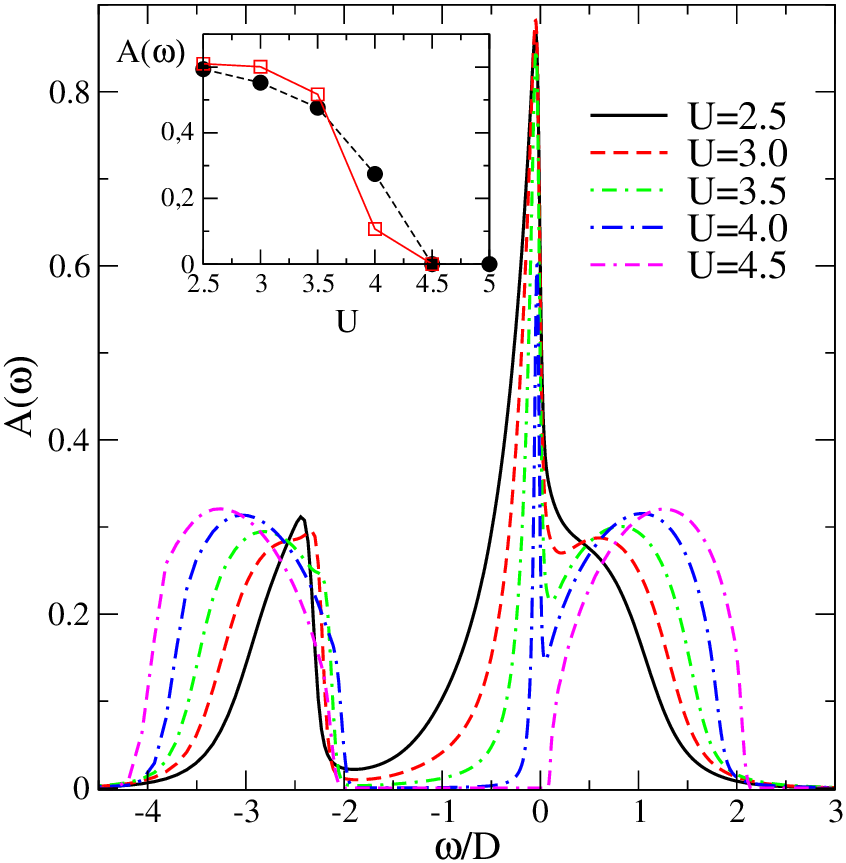}
\caption{(Color online) Spectral density $A(\omega)$ calculated by OCA for a non-symmetric case taking $2.5 < U < 5.0$ and an energy shift of -1.0 
from the corresponding symmetric case for each value of $U$. 
The inverse temperature is $\beta$ = 60. In the inset we show $A(\omega=0)$ as a function of $U$. 
The CT-QMC (OCA) data are displayed with the solid (dashed) lines and empty (filled) symbols, respectively. }
\label{Fig5}
\end{figure}

In Fig. \ref{Fig5}, the spectral densities calculated by OCA for different values of $U$ are shown. 
One may see that for the smallest value of $U$, the system is metallic with a large quasiparticle resonance that overlaps with the upper Hubbard band giving rise
to large charge fluctuations pertaining to a mixed valence regime. In the other extreme, for the largest value of $U$, the systems is an insulator with the Hubbard bands located 
symmetrically with respect to $\Delta\mu$. The value of the gap in this case is of the order of $2D$. In order to be able 
to describe accurately solutions with large gaps, we implemented a three-centered logarithmic mesh. 

By integrating $A(\omega)$ weigthed by the Fermi function for the corresponding temperature, we obtained the local occupancies in a very good agreement with the CT-QMC ones. 
It is not obvious that this quantity can be correctly evaluated within approximate analytical solvers. Hence, the fact that it is captured within OCA is important for the applicability of the method to non-symmetric cases.

In the inset of Fig.\ref{Fig5} we show $A(\omega=0)$ as a function of $U$ in comparison with CT-QMC. One sees that  both the OCA and CT-QMC indicate 
that the system turns an insulator for $U \geq 4.5$. For this level of doping there is no coexistence region and the OCA critical value $U_c$ agrees with 
the CT-QMC one within a 5\%.

Overall, we show that OCA also gives a very reasonable description of the Mott metal-insulating transition for the Hubbard model
away from half-filling.

\section{Summary and conclusions} \label{conclusions}
The self-consistent hybridization expansions in their different forms (NCA, OCA, symmetric finite-U NCA, etc) have been widely used not only
in the context of the impurity problem, but also in the framework of DMFT applied both to different lattice models 
and realistic cases, describing strongly correlated materials from first-principles. However, to the best of our knowledge,
a detailed and quantitative study of the Mott transition,  one of the essential problems of strongly correlated systems, 
has not been carried out up to now with these kind of approximate techniques. 

In this work, we asses the reliability of OCA impurity solver in the context of the DMFT method to describe the Mott metal-insulator
for the one band Hubbard model in the Bethe lattice
at half-filling within DMFT. We present the temperature-local repulsion $U$ phase diagram
in comparison with the numerically exact CT-QMC. We show that OCA can provide a 
very good quantitative description of the metal-insulator transition of the present model.
We obtain the metal-to-insulator  transition, $U_{c_2}$, within an error of less than 0.5\%  while the insulator-to-metal 
$U_{c_1}$ values are shifted to higher $U$ (about a 3\%) with respect to the CT-QMC one. 
We explain the overestimation of $U_{c_1}$ from an analysis of the second moment of 
the spectral density, $M_2$. We find that the expected theoretical value 
for $M_2$ is underestimated by OCA. 
Since $M_2$ is equal to the second moment of the spectral weight, we infer that the size of the gap in the insulating phase 
is also underestimated so that the Hubbard bands collapse for higher values of $U$ than for CT-QMC.

Aside from the Mott transition itself, we confirm previous results\cite{millis,ruegg} regarding the better performance of OCA 
in the insulating phase than in the metallic one. 
The high-frequency sum rules for the imaginary part of $\Sigma(i\omega)$ are obtained reasonably well in both phases, with the deviation in the 
insulating case being a bit smaller than in the metallic one. On the other hand, in the small frequency region the correlations are overestimated in 
the metallic case. This effect is also apparent in the value of the quasiparticle weigth that is underestimated by OCA, specially far away from the transition. 
This overestimation of the correlations in the metallic phase does not imply that OCA favors the insulating state, as has been previously 
stated in Ref. \cite{ruegg}, since we show the transition $U$ is well reproduced, especially the $U_{c_2}$ values. 
Furthermore, we show that the gap of the insulating phase is underestimated by OCA.

Finally, we study the perfomance of OCA for a non-symmetric case obtaining an overall reasonable agreement with CT-QMC, 
and a very similar critical value of $U$ for the Mott transition at the considered temperature. 
The study of non-symmetric cases are particularly relevant for applications to real materials.

Despite the above mentioned  deviations of OCA from exacts results, we are not aware of any other approximated technique yielding a phase-diagram with
this level of agreement with numerically-exact many-body methods.

\section{Acknowledgments}
This work was partially supported by CONICET, PIP 00273 and 01060 and MINCYT-ANPCyT, program ECOS-MINCyT France-Argentina (project A13E04), PICT 1875 and R1776, Argentina.

\end{document}